\documentclass[aps,prd,preprint,preprintnumbers,unsortedaddress,superscriptaddress,showpacs,nofootinbib]{revtex4-1}

\pdfoutput=1
\usepackage{bm}
\usepackage{graphicx}
\usepackage{amsmath}
\usepackage{amsfonts}
\usepackage{amssymb}
\usepackage{color}

\usepackage[section]{placeins}

\usepackage{booktabs}%改表格线条粗细hline, 使用\toprule, \midrule and \bottomrule
\usepackage[colorlinks, citecolor=blue,anchorcolor=red,menucolor=red, linkcolor=red,filecolor=red,runcolor=red,urlcolor=blue,frenchlinks=red]{hyperref}

\begin{document}%
%\date{\today}
\bibliographystyle{unsrt}
\arraycolsep1.5pt

\title{Triangle singularity in $\tau^- \to  \nu_\tau \pi^- f_0(980)$ ($a_0(980)$) decays}

\author{L.~R.~Dai}
\email{dailr@lnnu.edu.cn}
\affiliation{Department of Physics, Liaoning Normal University, Dalian 116029, China}
\affiliation{Departamento de F\'isica Te\'orica and IFIC, Centro Mixto Universidad de Valencia-CSIC,
Institutos de Investigac\'ion de Paterna, Aptdo. 22085, 46071 Valencia, Spain
}

\author{Q. X. Yu}
\email{yuqx@mail.bnu.edu.cn}
\affiliation{College of Nuclear Science and Technology, Beijing Normal University, Beijing 100875, China}
\affiliation{Departamento de F\'isica Te\'orica and IFIC, Centro Mixto Universidad de Valencia-CSIC,
Institutos de Investigac\'ion de Paterna, Aptdo. 22085, 46071 Valencia, Spain
}

\author{E.~Oset}
\email{oset@ific.uv.es}
\affiliation{Departamento de F\'isica Te\'orica and IFIC, Centro Mixto Universidad de Valencia-CSIC,
Institutos de Investigac\'ion de Paterna, Aptdo. 22085, 46071 Valencia, Spain
}

\date{\today}
\begin{abstract}
We study the triangle  mechanism  for the decay  $\tau^- \to  \nu_\tau \pi^- f_0(980)$,  with the
 $f_0(980)$ decaying  into $\pi^+ \pi^- $.  This process is initiated by $\tau^- \to \nu_\tau K^{*0} K^-$ followed by the
$K^{*0}$ decay into $\pi^- K^+$, then the $K^- K^+$  produce the $f_0(980)$ through a triangle loop containing $ K^* K^+ K^-$ which develops
 a singularity around $1420$~MeV in the $\pi f_0(980)$ invariant mass. We find a narrow peak in the $\pi^+ \pi^-$ invariant mass  distribution,
 which originates from the $f_0(980)$ amplitude.  Similarly, we also study the  triangle  mechanism  for the decay $\tau \to  \nu \pi^- a_0(980)$,  with the  $a_0(980)$
 decaying  into $\pi^0 \eta $.  The final branching ratios for $\pi^- f_0(980)$ and $\pi^- a_0(980)$ are of the order of $4 \times 10^{-4}$
 and $7 \times 10^{-5}$, respectively, which are within present measurable range. Experimental  verification of these predictions will shed light on the nature of the scalar
 mesons and on the origin for the ``$a_1(1420)$" peak observed in other reactions.
\end{abstract}

\maketitle

\section{Introduction}
\label{intro}
Triangle singularities were studied in detail by Landau  \cite{landau} and they emerge from a process symbolized by a triangle Feynman diagram  in which one particle
decays into $1$ and $R$, $R$ decays later into $2+3$ and  $1+2$  merge to give another state, or simply rescatter.  Under certain conditions where all particles $1,2,3$ can be
placed on shell, $1$ and $3$ are antiparallel and the process can occur  at the classical level \cite{coleman} (Coleman Norton Theorem), the process develops a singularity
visible in a peak in the corresponding cross sections. While no clear such physical  processes were observed for a long time, the situation  reverted recently where  clear cases have been
observed and  many reactions have been suggested  to show such phenomena. A particular case is the triangle singularity studied in \cite{ketzer,liu,aceti} where a peak seen by the
COMPASS collaboration in the $\pi f_0(980)$ final state \cite{compass}, branded originally  as a new resonance, ``$a_1(1420)$" , was naturally explained in terms of the triangle singularity stemming
from  the original production of $K^* \bar{K}$, decay of $K^*$ into $\pi K$ and fusion of $K\bar{K}$ to give the $f_0(980)$ resonance.

The interest in triangle singularities has grown recently. In addition to the interpretation of the   ``$a_1(1420)$" as a triangle singularity, the $f_1(1420)$, officially in the PDG tables \cite{pdg}
was also shown  to correspond to the ``$f_1(1285)$"  decay into $K^*\bar{K}$, with the ``$\pi a_0(980)$ decay width" \cite{barberis} also corresponding  to the ``$f_1(1285)$"  \cite{debastiani}.
Similarly the ``$f_2(1810)$"  was also  shown to come from  a triangle singularity \cite{xie}. Some particular  reactions have also been studied and partial contributions or peaks  in the cross
sections have also been associated  to triangle singularities, and suggestions of new reactions to see them have been proposed
\cite{dailam,szczepaniak,szczepaniak2,bondar,pilloni,pavao,liu2,sakairamos,rocaprc,daris,wuzou,acetiwu,wuwu,liuli,ewang,xieguo,caozhao,liangsa,desaos}.

In the present work we study the reactions  $\tau^- \to  \nu_\tau \pi^- f_0(980)$  and  $\tau^- \to  \nu_\tau \pi^- a_0(980)$.  The original $\tau^- $ decays  into a $\nu_\tau$ and a  $d\bar{u}$
state that has $I_3=-1, I=1$. The further hadronization including a $\bar{q}q$ pair  forms two mesons conserving isospin.  Hence, both decays modes are allowed. Since $f_0(980)$, $a_0(980)$  couple mostly
to $K\bar{K}$, the reaction requires the formation of this pair, in addition to the $\pi^-$. Hence it proceeds via $K^* \bar{K}$ production, followed by $K^*$ decay to $\pi^- K$
and the $K\bar{K}$ fuse to produce the  $f_0(980)$ or the $a_0(980)$. Then we have a triangle mechanism  that could or not produce a singularity. However we show that  it develops
a triangle singularity  at an invariant mass $M_{\rm inv} (\pi R) (R\equiv f_0, a_0) \simeq 1420$ MeV.  Interestingly, the triangle mechanism  that produces a peak  in this invariant mass
distribution is the same one that produced the ``$a_1(1420)$" peak observed in the COMPASS experiment.

The other issue present in this reaction is the $G$-parity.  The $\pi^- f_0(980)$ and $\pi^- a_0(980)$ have negative and positive $G$-parity respectively.  The formalism has to provide the means to filter the
states of $G$-parity just after the weak decay,  from the operators involved in the $W d\bar{u}$ vertex.  Fortunately a formalism has been developed recently \cite{tdai} in which the $G$-parity
appears explicitly in the amplitudes written at the macroscopic meson level after the hadronization to produce two mesons. By means of this  formalism  we can easily evaluate the loops involved  in the
 triangle mechanism  and predict quantitative mass distributions for the $\tau^-$ decay in these modes.  This is made possible because the radial matrix elements of the quark  wave functions, which
 are a source of large uncertainties  and we do not explicitly evaluate, are implicitly taken into account  by making use of the experimental value of the  $\tau \to \nu_\tau K^{*0} K^-$
 branching ratio, which is the first step in our loop  mechanism.

 By means of this approach we obtain  $\frac{d^2 \Gamma}{d M_{\text{inv}}(\pi^- R) d M_{\text{inv}}(\pi^+ \pi^-)}$  or $\frac{d^2 \Gamma}{d M_{\text{inv}}(\pi^- R) d M_{\text{inv}}(\pi^0 \eta)}$ which
 show the shapes of the  $f_0(980)$  and $a_0(980)$ resonances in the $\pi^+ \pi^- $ or $\pi^0 \eta$ mass distributions  respectively.   Then we integrate over the $\pi^+ \pi^- $ or $\pi^0 \eta$ invariant masses
 and obtain $\frac{d^2 \Gamma}{d M_{\text{inv}}(\pi^- R)}$, which shows a clear peak around $M_{\text{inv}}(\pi^- R) \simeq 1420$ MeV.  The further integration over $M_{\text{inv}}(\pi^- R)$ provides
 us branching ratios for $\tau^- \to  \nu_\tau \pi^- f_0(980)$  and  $\tau^- \to  \nu_\tau \pi^- a_0(980)$ production, and we obtain values  of $4 \times 10^{-4}$
 and $7 \times 10^{-5}$ for these two ratios  respectively, which are well within measurable range.

 The measurement of such reactions and comparison  with the present results should be very useful  since it conjugates  several interesting issues:
\begin{description}
  \item[i] It provides one more measurable example of a triangle singularity, which have been quite sparse up to now.
  \item[ii] It serves as a further test of the nature of the $f_0(980)$ and $a_0(980)$, since they are not directly produced from the weak decay, but come from fusion of $K\bar{K}$
  in a scattering process,  establishing a link with the chiral unitary approach to these resonances  where they are shown not to correspond to $q\bar{q}$ state but are generated by the scattering
  of pseudoscalar  mesons in coupled channels.
  \item[iii] The filters of the $G$-parity in the amplitudes can also provide information that can be extrapolated to $\tau^- \to  \nu_\tau M_1 M_2$ decays with $M_1 M_2$  pairs of states that have
  a given $G$-parity as $\pi \rho$, $\pi \omega$, $\eta \rho$ and $\eta' \rho$.
\end{description}

With the possible advent of a future $\tau^-$  facility, \footnote{Discussions are currently under way for such a facility in China (X. G. He, private communication)} predictions
like the present one and the motivation given, should provide the grounds for proposals at that machine. Yet, other existing  facilities have also access to these reactions since rates of
$10^{-5}$  and smaller are common in $\tau^-$ decays \cite{pdg}.

A reaction close to the present one is the $\tau^- \to  \nu_\tau \pi^- f_1(1285)$. The reaction  has been  measured \cite{pdg} with a branching ratio $(3.9 \pm 0.5)\times 10^{-4}$. In \cite{roca}
a mechanism  similar to the  present one is presented in which $K^{*} \bar{K}$  in an intermediate  state  merge to  produce the $f_1(1285)$, also dynamically generated  from the
 $K^{*} \bar{K}$  interaction \cite{roca2,geng}. In this case a triangle singularity  appearing around  $1800$ MeV in the  $\pi^- f_1(1285)$ invariant mass only shows up at the end of the
 phase space, such that  no visible peak associated to this triangle singularity  is seen in the mass distribution and other possible  interpretations are possible \cite{volkov}.  In the
 present case we shall see that the peak in the $\pi^- f_0 (a_0)$ mass distributions is very  strong and clear.

\section{Formalism}
\begin{figure}[ht]
\includegraphics[scale=0.6]{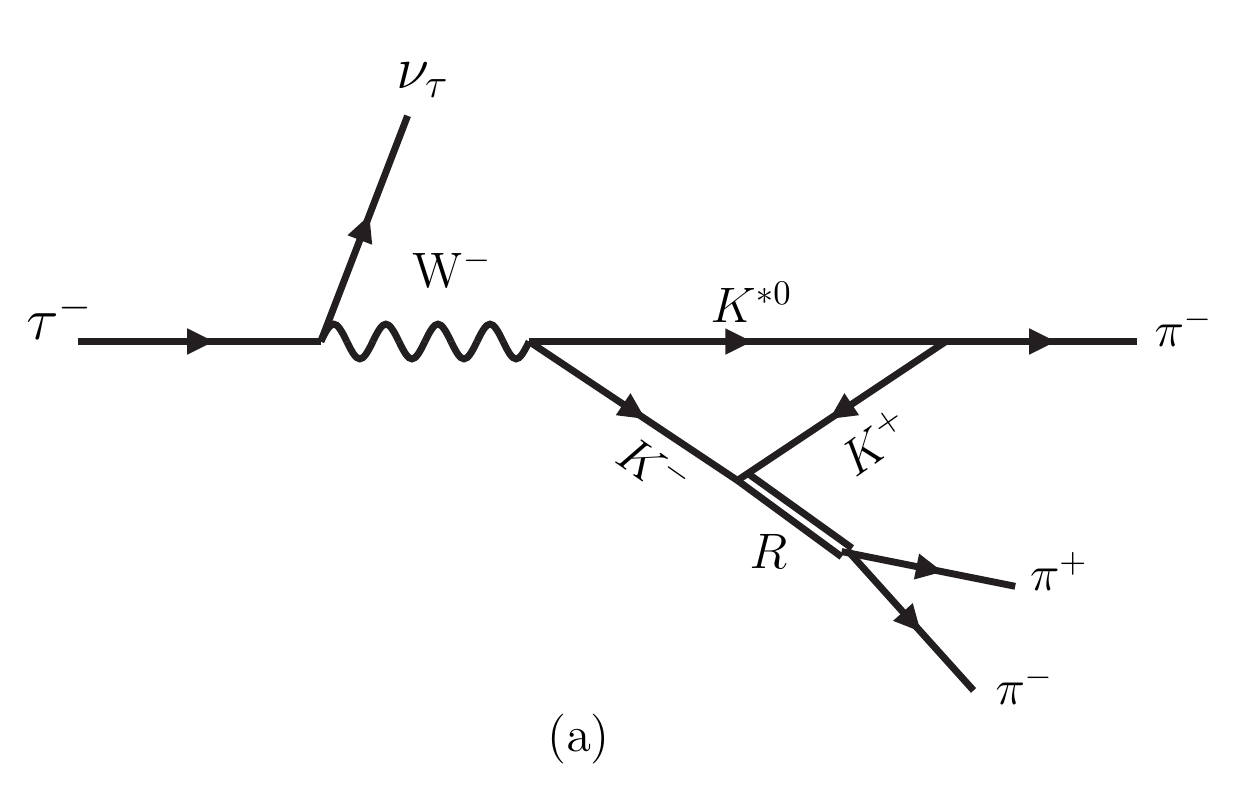} \includegraphics[scale=0.6]{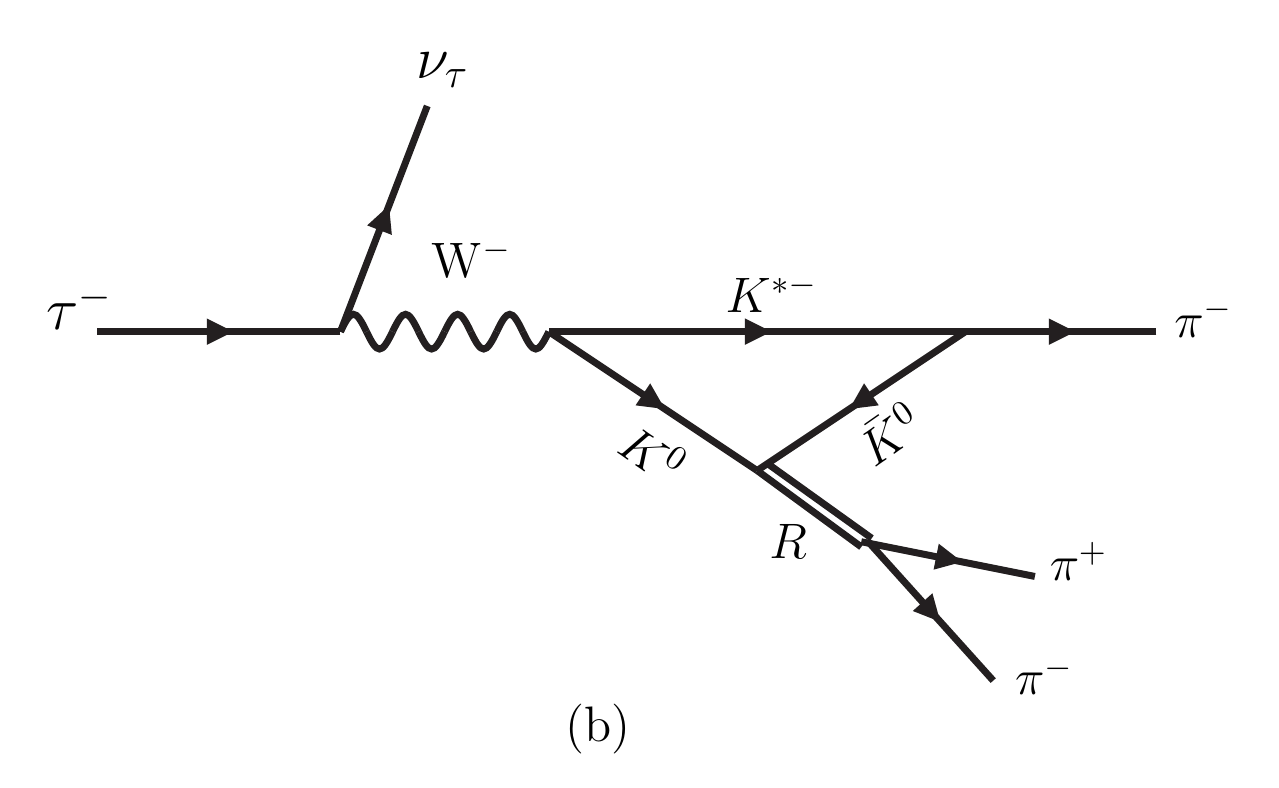}
\caption{ Diagram for the decay of  $\tau^- \to \nu_\tau \pi^- \pi^+ \pi^- $.  (a) The process $\tau^- \to \nu_\tau K^{*0} K^-$ followed by the
$K^{*0}$ decay into $\pi^- K^+$  and the merging of the $K^- K^+$ into  $f_0(980)$;
(b) The process $\tau^- \to \nu_\tau K^{*-} K^0$  followed by the $K^{*-} $ decay into $\pi^-\bar{K}^0$ and the merging of the $ K^0 \bar{K}^0$ into $f_0(980)$.}
\label{fig:diaf0}
\end{figure}

\begin{figure}[ht]
\includegraphics[scale=0.6]{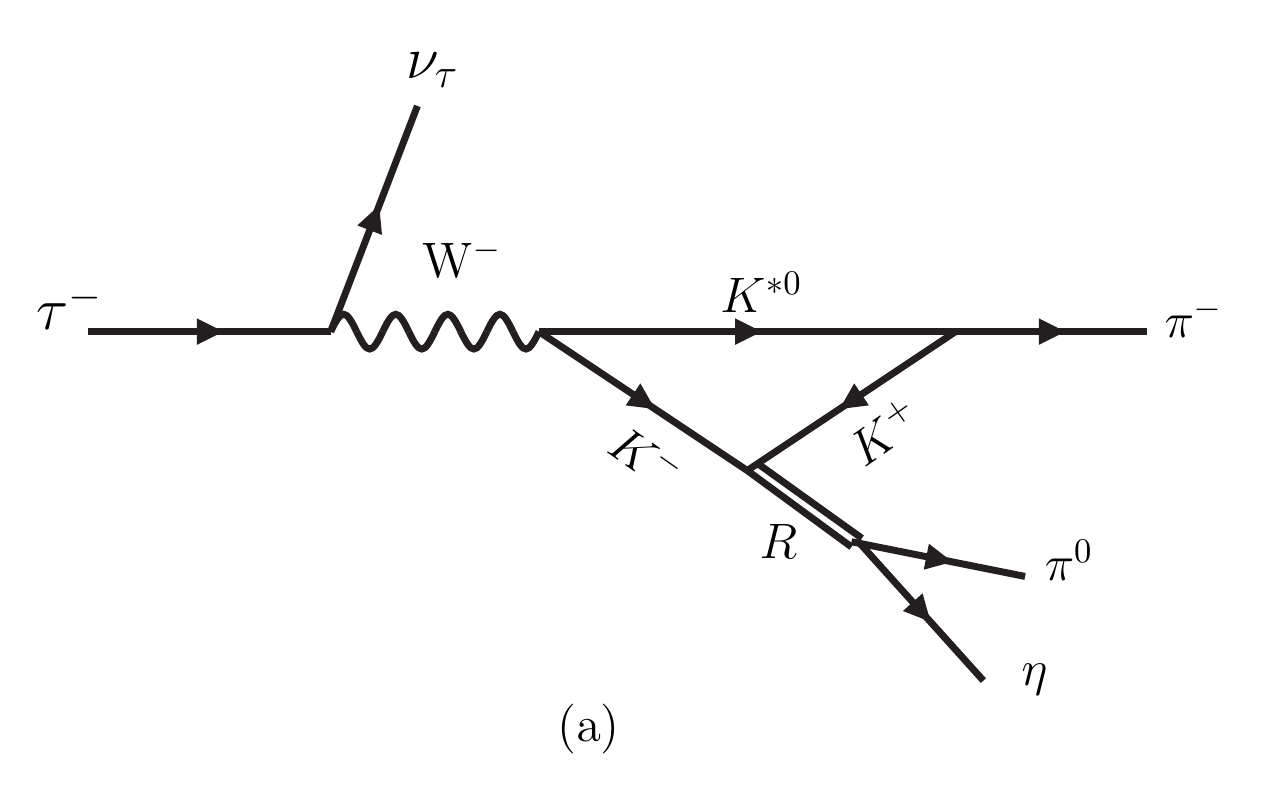} \includegraphics[scale=0.6]{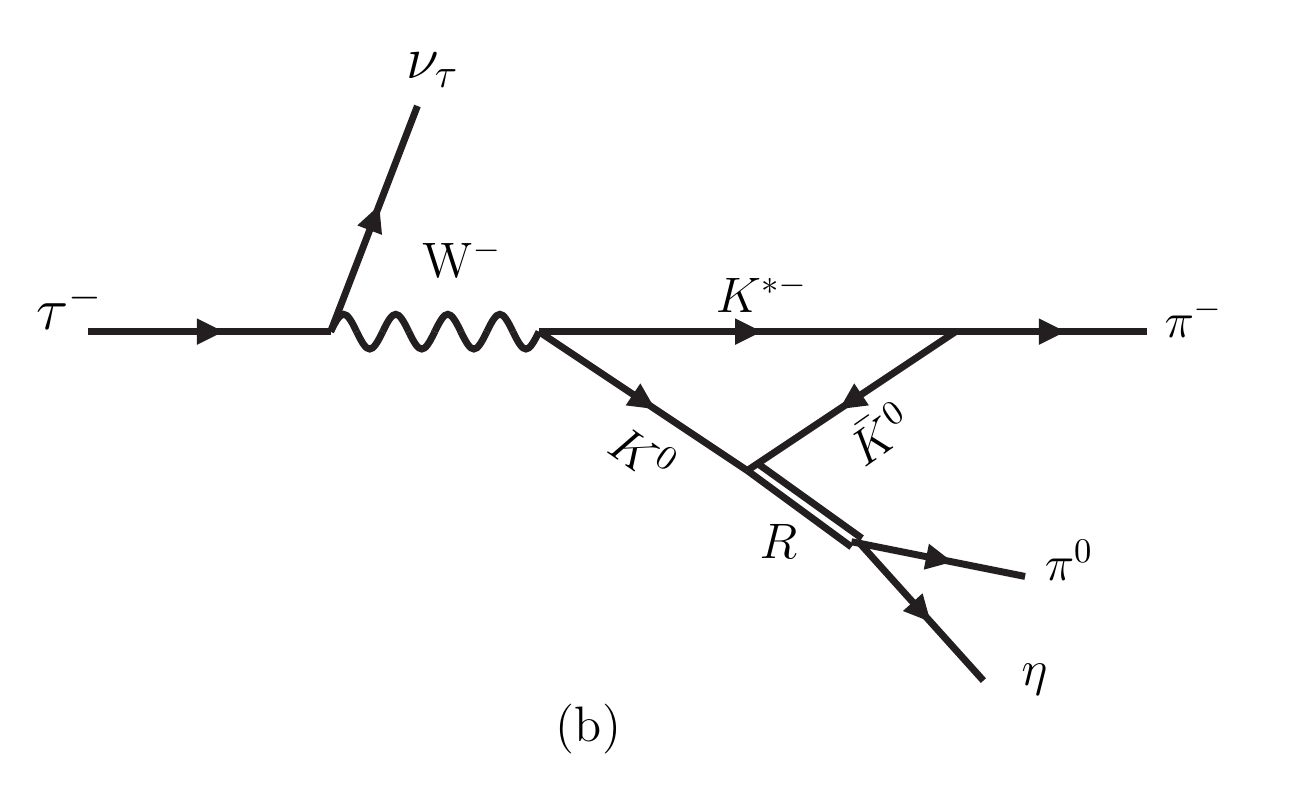}
\caption{ Diagram for the decay of  $\tau^- \to \nu_\tau \pi^- \pi^0 \eta $.  (a) The process $\tau^- \to \nu_\tau K^{*0} K^-$ followed by the
$K^{*0}$ decay into $\pi^- K^+$  and the merging of the $K^- K^+$ into  $a_0(980)$;
(b) The process $\tau^- \to \nu_\tau K^{*-} K^0$  followed by the ${K}^{*-} $ decay into $\pi^-\bar{K}^0$ and the merging of the $ K^0 \bar{K}^0$ into $a_0(980)$.}
\label{fig:diaa0}
\end{figure}

We will study the effect of triangle singularities in the decay of  $\tau^- \to \nu_\tau \pi^- \pi^+ \pi^- $  and  $\tau^- \to \nu_\tau \pi^- \pi^0 \eta $ decays with $\pi^+ \pi^-$ forming the
$f_0(980)$ and  $\pi^0 \eta$ the $a_0(980)$.
 The complete Feynman diagrams for the decay with the triangle mechanism through the $f_0(980)$ and  $a_0(980)$
 are shown in Figs. \ref{fig:diaf0} and \ref{fig:diaa0}.

In Fig.\ref{fig:diaf0}, we investigate the $\tau^- \to \nu_\tau \pi^- \pi^+ \pi^- $ decay
via $f_0(980)$ formation, where Fig. \ref{fig:diaf0}(a) shows the process $\tau^- \to \nu_\tau K^{*0} K^-$ followed by the
$K^{*0}$ decay into $\pi^- K^+$  and the merging of the $K^- K^+$ into  $f_0(980)$, and Fig.\ref{fig:diaf0}(b) shows  the process $\tau^- \to \nu_\tau K^{*-} K^0$
followed by the $\bar{K}^{*0} $ decay into $\pi^-\bar{K}^0$ and the merging of the $ K^0 \bar{K}^0$ into $f_0(980)$.
Each process generates a singularity, and we will see a signal for the isospin $I=0$ resonance state $f_0(980)$ formation in the invariant mass of $\pi^+ \pi^-$.
In the  study of Refs.~\cite{oller,nieves,kaiser,daiplb,locher}, the $f_0(980)$  appears as the
dynamically generated state from the  $\pi^+ \pi^+$, $\pi^0 \pi^0$, $K^+ K^-$, $K^0 \bar{K}^0$, and $\eta\eta$  in the coupled-channels calculation.

\begin{figure}
	\begin{center}
		\includegraphics[width=0.45\textwidth]{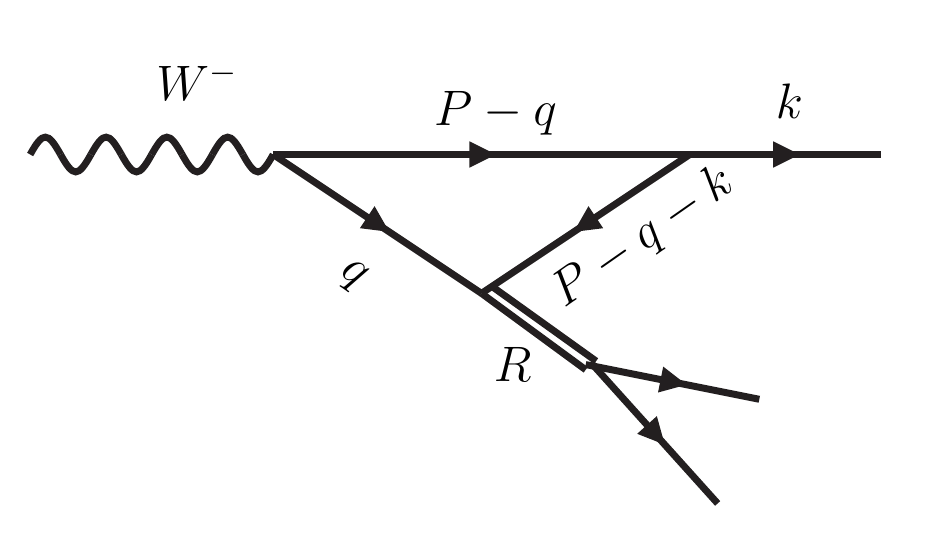}
	\end{center}
	\caption{\label{fig:mom} The momenta assignment for the decay process}
\end{figure}

Similarly,  in Fig. \ref{fig:diaa0},  we investigate the $\tau^- \to \nu_\tau \pi^- \pi^0 \eta$ decay via $a_0(980)$ formation,
where Fig. \ref{fig:diaa0}(a) shows the process $\tau^- \to \nu_\tau K^{*0} K^-$ followed by the $K^{*0}$ decay into $\pi^- K^+$  and the merging of the $K^- K^+$ into  $a_0(980)$,
and the process $\tau^- \to \nu_\tau K^{*-} K^0$ followed by the ${K}^{*-} $ decay into $\pi^-\bar{K}^0$ and the merging of the $ K^0 \bar{K}^0$ into $a_0(980)$.
Both processes  also generate a singularity, and we will see a signal for the isospin $I=1$ resonance  state $a_0(980)$  in the invariant mass of $\pi^0 \eta$.
In the study of Refs.~\cite{oller,nieves,kaiser,daiplb,locher}, the $a_0(980)$  appears as the dynamically generated state of  $K^+ K^-$, $K^0 \bar{K}^0$, and  $\pi^0 \eta$
in the coupled-channels calculation.  The momenta assignment for the decay process is given in Fig.~\ref{fig:mom}.

Let us address, next, the evaluation of the  $\tau \to \nu_\tau K^{*0} K^- ,\nu_\tau K^{*-} K^0$   parts.
The production is assumed to proceed first from the Cabibbo favored $\bar u d$ production from the $W^-$ which
then hadronizes producing an $s \bar s$ with quantum numbers of the vacuum, which are implemented with the $^3P_0$ model \cite{micu,oliver,bijker}.
This leads to the $K^{*0} K^-$  and $K^0 K^{*-}$  states with the same weight.
In  Ref.~\cite{tdai} the mechanism for hadronization is done in detail. The first step corresponds to the flavor combinations in the hadronization. There it is shown that $d(\bar{s}s)\bar{u}=(d\bar{s})s\bar{u}$
gives rise to $K^0 K^{*-}$ and $K^{*0} K^-$ with the same weight (see Eqs. (2) and (3) of  Ref.~\cite{tdai}).
 The second  step corresponds to the detailed study  of  the spin-angular momentum algebra to combine the quarks for the
 $^3P_0$ $\bar{s}s$ state ($L'=1, S'=1,J'=0$) with a $\bar{d}$  quark in $L=1$  to have finally $s$-wave production of the two mesons.  In  Ref.~\cite{tdai} the $p$-wave
vector-pseudoscalar production was ruled out based on the theoretical results, and experimental results that show the vector-pseudoscalar pairs coupling to axial
vector resonance $J^{PC}=1^{++} $\cite{pr88}, which proceeds with $s$-wave.
The needed results from \cite{tdai}  are given in the next subsection.
\subsection{$\tau \to \nu_\tau K^{*0} K^-$ decay }
The elementary quark interaction  is given by
\begin{eqnarray}\label{eq:c}
H= \mathcal{C} L^\mu Q_\mu \,,
\end{eqnarray}
where $\mathcal{C}$ contains the couplings of the weak  interaction.  The  leptonic current is given by
\begin{eqnarray}
L^\mu=\langle {\bar u}_\nu |\gamma^\mu-\gamma^\mu\gamma_5| u_\tau\rangle \,,
\end{eqnarray}
and  the quark current  by
\begin{eqnarray}
Q^\mu=\langle \bar u_d|\gamma^\mu-\gamma^\mu\gamma_5|v_{\bar u}\rangle  \,.
\end{eqnarray}
As is usual in the evaluation of decay widths to three final particles, we evaluate the matrix elements in the frame where the two mesons system is at rest.
For the evaluation of the matrix element $Q_\mu$ we assume that the quark spinors  are at rest in that frame \cite{tdai}, then we have $\gamma^0 \to 1$,  $\gamma^i \gamma_5 \to \sigma^i$
in terms of bispinors $\chi$ and  after the spin angular momentum combination we  have
\begin{eqnarray} \label{eq:Qu2}
Q_0&=& \langle \chi^{\prime} | 1 | \chi \rangle \to  M_0  \nonumber \, , \\
Q_i&=&  \langle \chi^{\prime} | \sigma_i |\chi \rangle \to  N_i \, .
\end{eqnarray}
Denoting for simplicity,
\begin{equation}\label{eq:new}
\overline{L}^{\mu\nu}= \overline{\sum} \sum  L^\mu {L^\nu}^\dagger  \, ,
\end{equation}
to obtain the $\tau$ width we  must evaluate
\begin{eqnarray}\label{eq:L}
\overline{\sum} \sum \left|t\right|^2 &=&\overline{\sum} \sum  L^\mu {L^\nu}^\dagger Q_\mu Q_\nu^*  \nonumber \, , \\
&=& \bar{L}^{00}\, M_0~ M^*_0
+\bar{L}^{0i}\,M_0 ~N^*_i
+\bar{L}^{i0} \, N_i ~M^*_0
+\bar{L}^{ij} N_i ~N_j^* \, ,
\end{eqnarray}
with $\overline{L}^{\mu\nu}$ given by
\begin{eqnarray}\label{eq:LL}
\overline{\sum} \sum  L^\mu {L^\nu}^\dagger =\frac{1}{m_{\nu} m_{\tau}}\left( p'^\mu p^\nu  +p'^\nu p^\mu - g^{\mu\nu}p'\cdot p+i \epsilon^{\alpha\mu\beta\nu}p'_\alpha p_\beta\right) \, ,
\end{eqnarray}
where $p,p'$ are the momenta of the $\tau$ and $\nu_\tau$ respectively and  we use the field normalization for fermions of Ref. \cite{mandl}.

From the work  \cite{tdai}  we obtain  the results for the  $J=1, J'=0$ case, which corresponds to the  $\tau \to \nu_\tau K^{*0} K^-$ decay.
\begin{eqnarray}\label{eq:Q1}
M_0 &=&\frac{1}{\sqrt{6}}\frac{1}{4\pi} \,,~~  ~~~ {\rm for ~ any ~~} M   \nonumber \, , \\
N_\mu &=& (-1)^{-\mu} \frac{1}{\sqrt{3}}\frac{1}{4\pi} {\cal C}(1 1 1; M,-\mu,M-\mu)  \, ,
\end{eqnarray}
where $M$ is the third component of $J$ and $\mu$ is the index of $N_i$ in spherical basis, with ${\cal C}(\cdots)$ a Clebsch-Gordan coefficient.

It was shown in \cite{tdai} that the order in which the vector and  pseudoscalar mesons are produced is essential to understand the $G$-parity symmetry  of these reactions. Then from \cite{tdai}
we write here the results for $PV$ production $J=0, J'=1$, which  corresponds to the $\tau \to \nu_\tau K^0 K^{*-}$ decay,
\begin{eqnarray}\label{eq:Q2}
M_0 &=&\frac{1}{\sqrt{6}}\frac{1}{4\pi} \,,~~  ~~~ {\rm for ~ any ~~} M'   \nonumber \, , \\
N_\mu &=& -(-1)^{-\mu} \frac{1}{\sqrt{3}}\frac{1}{4\pi} {\cal C}(1 1 1; M',-\mu,M'-\mu) \, .
\end{eqnarray}
 Note that while $M_0$  is the same for $VP$ and $PV$ productions, $N_i$ changes sign for $VP$ and $PV$. This sign is essential for the conservation of $G$-parity  in the reaction, as we shall see.
Indeed, at the quark  level  the primary $d\bar{u}$  state produced has $I_3=-1$ and hence $I=1$.  The $G$-parity of a $q\bar{q}$  pair is given by $(-1)^{L+S+I}$. As we mentioned $L=1, I=1$ and the spin
of the state is $0$ for the $1$ operator and $1$ for the $\sigma^i$ operator  of Eq. \eqref{eq:Qu2}. This means that the term $M_0$ proceeds with $G$-parity positive, while  $N_i$ has $G$-parity
negative. Since $\pi$, $f_0(980)$, and $a_0(980)$ have $G$-parity $-,+,-$ respectively, then $\pi^- f_0(980)$ will proceed with the $N_i$ amplitude,  while $\pi^- a_0(980)$  proceeds with
the $M_0$ term  and there is no simultaneous  contribution of the two terms in these reactions. This we shall  see analytically   when evaluating explicitly  the amplitudes for the processes of
Figs. \ref{fig:diaf0} and \ref{fig:diaa0}.

As seen in Eq. \eqref{eq:c}, we have  the unknown constant $\cal{C}$ in our approach which includes factors involving the matrix elements of the radial quark wave functions (the
spin-angular momentum variables are explicitly accounted for in the work of \cite{tdai}).  We then determine $\cal{C}$  from the experimental ratio of $\tau \to \nu_\tau K^{*0}K^-$.
For this we use the results of  \cite{tdai} for this reaction.

By taking the quantization axis along the direction of the neutrino in the $\tau^-$ rest frame, we find
\begin{eqnarray}\label{eq:ff2}
\overline{\sum} \sum \left|t\right|^2 &=&  \frac{{\cal{C}}^2}{m_\tau m_\nu} \left(\frac{1}{4\pi}\right)^2
 \left[\left(E_\tau E_\nu + {p}^2 \right) \frac{1}{2}{h}^2_i+ \left(E_\tau E_\nu -\frac{1}{3}  {p}^2  \right) \overline{h}^2_i \right]  \nonumber\, \\
 &=&  \frac{{\cal{C}}^2}{m_\tau m_\nu} \left(\frac{1}{4\pi}\right)^2
 \left(\frac{3}{2} E_\tau E_\nu +\frac{1}{6}{p}^2 \right)
 \end{eqnarray}
where  ${h}_i=\overline{h}_i=1$,  $p$ is the momentum of the $\tau$, or $\nu_\tau$, in the $K^{*0}K^- $ rest frame, given by
\begin{equation}\label{eq:newlabel}
p=p_\nu=p_\tau=\frac{\lambda^{1/2}(m^2_\tau,m^2_\nu,M_{\rm inv}^{2} (K^{*0} K^-))}{2 M_{\rm inv}{(K^{*0} K^-)}}\, ,
\end{equation}
and $E_\nu=p$, $E_\tau=\sqrt{m_\tau^2+p^2}$.

Now for $\tau \to \nu_\tau K^{*0}K^- $  decay, we obtain
\begin{equation}\label{eq:dGdM}
\frac{ d\Gamma}{d M_{\rm inv}{(K^{*0} K^-)}} =  \frac{2\,m_\tau 2\, m_\nu}{(2\pi)^3} \frac{1}{4 m^2_\tau}\, p'_\nu {\widetilde p_K}\, \overline{\sum} \sum \left|t\right|^2 \,,
\end{equation}
where $p'_\nu$ is the neutrino momentum in the $\tau$ rest frame
\begin{equation}
p'_\nu=\frac{\lambda^{1/2}(m^2_\tau,m^2_\nu,M_{\rm inv}^{2} (K^{*0} K^-))}{2 m_\tau}\, ,
\end{equation}
and  ${\widetilde p_K}$ the momentum of $K^{-}$  in the $K^{*0}K^-$ rest frame  given by
\begin{equation} \label{eq:new2}
\widetilde{p}_K=\frac{\lambda^{1/2}(M_{\rm inv}^{2} (K^{*0} K^-), m_{K^{*0}}^2, m_{K^-}^2)}{2 M_{\rm inv}{(K^{*0} K^-)}}\, .
\end{equation}

Experimentally,  the branching ratio of ${\cal B} (\tau \to \nu_\tau K^{*0} K^-)$ decay,
\begin{eqnarray}\label{eq:Ce}
{\cal B} (\tau \to \nu_\tau K^{*0} K^-) = \frac{1}{\Gamma_\tau} \Gamma(\tau \to \nu_\tau K^{*0} K^-) =(2.1 \pm 0.4)\times 10^{-3},
\end{eqnarray}
and then
\begin{eqnarray}\label{eq:C}
 \frac{{\cal C}^2}{\Gamma_\tau} = \frac{{\cal B} (\tau \to \nu_\tau K^{*0} K^-)}{ \int_{m_{K^-}+m_{K^{*0}}}^{m_{\tau}}\frac{1}{(2\pi)^3} \frac{1}{m^2_\tau} p'_\nu {\widetilde p_K}\frac{1}{(4\pi)^2}\left(\frac{3}{2} E_\tau E_\nu +\frac{1}{6} \,{p}^2 \right)dM_{\rm inv}{(K^{*0} K^-)} }  \,  ,
\end{eqnarray}
from which  we can evaluate the value of the constant ${\cal C}^2$.

\subsection{Evaluation of  the triangle diagram}
In Eq. \eqref{eq:Q1} we need $M$, the third component of $J$. In order to evaluate the loops of Figs.\ref{fig:diaf0}, \ref{fig:diaa0}, we find most convenient to take the $z$ direction  along the momentum $\bm{k}$ of the pion produced (see Fig.\ref{fig:mom}). Indeed, in the $\pi f_0(980)$ rest frame, where we evaluate  the amplitude, ${\bm P}=0$.  The vertex $K^* \to K \pi$ is of the type ${\bm\epsilon}\cdot ({\bm k}+{\bm q}+{\bm k})$
\footnote{Since in the triangle singularity the the $K^{*0} K^-$ intermediate states are placed on shell, and have a small momentum compared to the $K^{*}$ mass, we neglect the ${\bm\epsilon}^0$ component,  which was found in \cite{sakairamos} to be an excellent approximation in such a case.}. The ${\bm q}$ integration of $ \int d^3q~ q_i~ k_i \cdots \cdots $ will necessarily give something proportional to $ {\bm k}$, which is the only
non integrated vector in the loop integral. Hence, we have an effective vertex of the type ${\bm\epsilon}\cdot {\bm k}$. If the $z$ direction is chosen along ${\bm k}$, this  selects only the ${\bm\epsilon}_z$ component (${\bm\epsilon}_0$ in spherical basis) and  ${\bm\epsilon} \cdot {\bm k}=|{\bm k}|=k$.  This also means that only $M=0$ contributes in the loop and this allows us to calculate  trivially the $M_0$, $N_\mu$ amplitude in that
frame. Indeed for $J=1, J'=0$,
\begin{eqnarray}\label{eq:Q3}
M_0 &\to& \frac{1}{\sqrt{6}}\frac{1}{4\pi}    \nonumber \, , \\
N_\mu &\to&  (-1)^{-\mu} \frac{1}{\sqrt{3}}\frac{1}{4\pi} {\cal C}(1 1 1; 0,-\mu,-\mu) \, ,
\end{eqnarray}
and for $J=0, J'=1$, $M_0$ is the same  and $N_\mu$ changes sign.

Explicit calculation of the  Clebsch-Gordan coefficients in Eq.\eqref{eq:Q3} gives
\begin{eqnarray}
N_{\mu=+1} =-\frac{1}{\sqrt{3}} \frac{1}{4 \pi} \frac{1}{\sqrt{2}} \, ,\qquad N_{\mu=-1} =\frac{1}{\sqrt{3}} \frac{1}{4 \pi} \frac{1}{\sqrt{2}} \, ,\qquad N_{\mu=0}=0 \, ,
\end{eqnarray}
which in cartesian coordinate can be written  as
\begin{eqnarray}
N_{i}=\frac{1}{\sqrt{3}} \frac{1}{4 \pi} \delta_{i 1} \, ,
\end{eqnarray}
the index $1$ for the $x$ direction. We now define the triangle loop functions, $t_L$, such that
\begin{eqnarray}
t_L g t_{K^+ K^-,\pi^+ \pi^-} {\bm k} &=& i \int \frac{d^4 q}{(2 \pi)^4}   \frac{1}{q^2-m^2_{k^-}+i \epsilon} \frac{1}{(P-q)^2-m^2_{K^{*0}}+i \epsilon} \nonumber \, , \\
& \times &  \frac{1}{(P-q-k)^2 - m_{K^+}^2+i \epsilon} \, g\,   (2 {\bm k} + {\bm q}) \,  t_{K^+ K^-,\pi^+ \pi^-} \, ,
\label{eq:tt}
\end{eqnarray}
where the $K^{*0} \to \pi^- K^+$ vertex has been evaluated from the $VPP$ Lagrangian
\begin{equation}
\mathcal{L}_{VPP} = -i g \left < V^{\mu} \left[P, \partial_{\mu} P\right] \right >  \, ,
\label{eq:vpp}
\end{equation}
and the brackets $\left<...\right>$ mean the trace over the SU(3) flavour matrices,  with the coupling $g$ given by $g=m_V/2 f_{\pi}$ in the local hidden
gauge approach, with $m_V=800 \ \text{MeV}$ and $f_{\pi}$=93 MeV.

As mentioned above,
\begin{eqnarray}
\int f({\bm k},{\bm q}) q^i =A k^i \, , \qquad  A=\int f({\bm k},{\bm q}) \frac{\bm{q} \cdot \bm{k}}{|\bm{k}|^2}\, ,
\end{eqnarray}
Hence, ${\bm q}$ in $ 2 {\bm k}+{\bm q}$ in Eq. \eqref{eq:tt} can be replaced effectively by $\frac{\bm{q} \cdot \bm{k}}{|\bm{k}|^2} {\bm k} $.  By performing
analytically the $q^0$ integration in Eq. \eqref{eq:tt} we find \cite{aceti2,guo}
\begin{widetext}
\begin{align}\label{eq:tt2}
 t_T =& \int \frac{d^3 q}{(2 \pi)^3} \frac{1}{8 \omega_{K^{*}} \omega_{K^+} \omega_{K^-}} \frac{1}{k^0-\omega_{K^+}-\omega_{K^{*}}+i\frac{\Gamma_{K^{*}}}{2}} \frac{1}{P^0+\omega_{K^-}+\omega_{K^+}-k^0} \left(2+ \frac{\bm{q} \cdot \bm{k}}{|\bm{k}|^2}\right) \nonumber \,  \\
&\times \frac{1}{P^0 - \omega_{K^-} -\omega_{K^+}-k^0+ i \epsilon}  \frac{2P^0 \omega_{K^-} + 2 k^0 \omega_{K^+} -2(\omega_{K^-}+\omega_{K^+})(\omega_{K^-}+\omega_{K^+}+\omega_{K^{*}})}{P^0-\omega_{K^{*}}-\omega_{K^-}+i\frac{\Gamma_{K^{*}}}{2}},
\end{align}
\end{widetext}
with  $P^0=M_{\rm inv}(\pi^- f_0)$,  $\omega_{K^-}=\sqrt{\bm{q}^2+m_K^2}$, $\omega_{K^+}=\sqrt{({\bm q}+{\bm k})^2+m_{K}^2}$, and $\omega_{K^*}=\sqrt{\bm{q}^2+m_{K^\ast}^2}$
\begin{equation}
k^0=\frac{M^2_{\rm inv}(\pi^- f_0)+m_{\pi}^2-M^2_{\rm inv}(\pi^+ \pi^-)}{2 M_{\rm inv}(\pi^- f_0)},
\end{equation}
\begin{equation} \label{eq:25}
k=\frac{\lambda^{1/2}(M^2_{\rm inv}(\pi^- f_0), m_{\pi}^2, M^2_{\rm inv}(\pi^+ \pi^-))}{2 M_{\rm inv}(\pi^- f_0)}.
\end{equation}
Similarly, we can get the  triangle amplitude for the $\pi^- a_0$ case. Note also that an $i \epsilon$ in the propagators involving  $\omega_{K^{*}}$ is  replaced by $i\frac{\Gamma_{K^{*}}}{2}$.

Then the formalism for the loop diagrams can be done as for the  $K^{*0}K^- $ production replacing
\begin{eqnarray}
M_0 \to  \widetilde{M}_0   t_{K^+ K^-, \pi^+ \pi^-} \,; \quad  \widetilde{M}_0 = g \frac{1}{\sqrt{6}}\frac{1}{4\pi} \,k \,t_L  \nonumber \, , \\
N_i \to  \widetilde{N}_i   t_{K^+ K^-, \pi^+ \pi^-} \,; \quad   \widetilde{N}_i= g \frac{1}{\sqrt{3}}\frac{1}{4\pi} \,k \,t_L \, \delta_{i1} \,
\end{eqnarray}
and  for $K^0 K^{*-}$,  $\widetilde{M}_0$ is the same and  $\widetilde{N}_i$ changes sign.

The combination of the diagram of Fig. \ref{fig:diaf0}(b) proceeds in a similar way. The changes are:  $t_{K^+ K^- \to \pi^+ \pi^-}$ is replaced by $t_{K^0 \bar{K}^0 \to \pi^+ \pi^-}$
and the $K^{*-} \to \pi^- \bar{K}^0$ vertex  has opposite sign to $K^{*0} \to \pi^- K^+$. Then, the sum of the two terms is taken into account by means of
\begin{eqnarray}\label{eq:M0}
M_0  &\to&  \widetilde{M}_0 (K^{*0} K^-) \, t_{K^+ K^-,\pi^+ \pi^-} - \widetilde{M}_0 (K^{*-} K^0) \, t_{K^0 \bar{K}^0,\pi^+ \pi^-}             \nonumber \,  \\
&=& \widetilde{M}_0 (K^{*0} K^-) \left(t_{K^+ K^-,\pi^+ \pi^-} -t_{K^0 \bar{K}^0,\pi^+ \pi^-}\right) \, ,
\end{eqnarray}
\begin{eqnarray}\label{eq:Ni}
N_i  &\to&  \widetilde{N}_i (K^{*0} K^-) \, t_{K^+ K^-,\pi^+ \pi^-} - \widetilde{N}_i (K^{*-} K^0) \, t_{K^0 \bar{K}^0,\pi^+ \pi^-}             \nonumber \,  \\
&=& \widetilde{N}_i (K^{*0} K^-) \left(t_{K^+ K^-,\pi^+ \pi^-} + t_{K^0 \bar{K}^0,\pi^+ \pi^-}\right) \, .
\end{eqnarray}
When we have $\pi^0 \eta$ production, as in Fig. \ref{fig:diaa0}, the formalism is identical, we only replace  $\pi^+ \pi^-$ by $\pi^0 \eta$ at the end in $t_{K \bar{K} \to m'_1 m'_2}$.
Next, in order to have isospin conservation and hence proper $G$-parity state we will solve the $t_{ m_1 m_2 \to m'_1 m'_2}$ amplitudes with average masses for the kaons and
 average masses for the pions and we shall also take  average masses  for $K^*$ masses in the loop. In this case we have
\begin{eqnarray}
t_{K^+ K^-, \pi^+ \pi^-} &=& t_{K^0 \bar{K}^0, \pi^+ \pi^-} \nonumber \, , \\
t_{K^+ K^-, \pi^0 \eta~} &=& - t_{K^0 \bar{K}^0,\pi^0 \eta~}  \, .
\end{eqnarray}
Hence in the case of the amplitude $M_0$ in Eq. \eqref{eq:M0} and $\pi^+ \pi^-$ in the final state we find a cancellation of the amplitudes for  diagram of Figs. \ref{fig:diaf0} (a) and \ref{fig:diaf0} (b).
If instead we have $\pi^0 \eta$ in the end, the two diagrams of  Figs. \ref{fig:diaa0} (a) and \ref{fig:diaa0} (b) give the same contribution and  sum coherently. Conversely, in the
$N_i$ term of Eq. \eqref{eq:Ni} the two terms corresponding to  Figs. \ref{fig:diaf0} (a) and \ref{fig:diaf0} (b) add and those  of Figs. \ref{fig:diaa0} (a) and \ref{fig:diaa0} (b) cancel exactly.
In summary, the  $M_0$ terms cancel for the production of $f_0(980)$ and add for the production of $a_0(980)$.  This is, the $f_0(980)$ production proceeds via the $N_i$ term
and the $a_0(980)$  production  via the $M_0$ term. Since $\pi^- f_0(980)$ has negative $G$-parity  and $\pi^- a_0(980)$  positive $G$-parity, we confirm that  the $M_0$ term
in the loop   corresponds to positive $G$-parity and the $N_i$ term  to negative $G$-parity, as we found earlier at the quark level.

Then for $\pi^- f_0(980)$ we will have
\begin{eqnarray} \label{eq:tf0}
\overline{\sum} \sum \left|t\right|^2 &=& \bar{L}^{ij} \widetilde{N}_i ~\widetilde{N}_j^* \, g^2\, |\,2 \,t_{K^+ K^-,\pi^+ \pi^-}|^2   \nonumber \, , \\
&=& \frac{{\cal C}^2}{m_\tau m_\nu}  \left(E_\tau E_\nu -\frac{1}{3} p^2 \right)   \frac{1}{3} \frac{1}{(4 \pi)^2} \,k^2 |t_L|^2  \, g^2\, |\,2 \,t_{K^+ K^-,\pi^+ \pi^-}|^2  \, .
\end{eqnarray}

Similarly, for the production of $\pi^- a_0(980)$ we will have
\begin{eqnarray}\label{eq:ta0}
\overline{\sum} \sum \left|t\right|^2 &=& \bar{L}^{00} \widetilde{M}_0 ~\widetilde{M}_0^* \, g^2\, |\,2 \,t_{K^+ K^-,\pi^0 \eta}|^2   \nonumber \, , \\
&=& \frac{{\cal C}^2}{m_\tau m_\nu}  \left(E_\tau E_\nu + p^2 \right)   \frac{1}{6} \frac{1}{(4 \pi)^2} \, k^2 |t_L|^2 \, g^2\, |\,2 \,t_{K^+ K^-,\pi^0 \eta}|^2   \, .
\end{eqnarray}
where we have taken into account that $p_i \delta_{i1} p_j \delta_{j1}$ is $p^2_x$ and when integrated  over the phase space gives rise to $\frac{1}{3} p^2$.

For  $\tau^- \to \nu_\tau \pi^- \pi^+ \pi^- $ decay,  the double  differential  mass distribution for $M_{\rm inv}(\pi^+ \pi^-)$  and $M_{\rm inv}(\pi^- f_0)$  is given by \cite{pavao}
\begin{eqnarray}\label{eq:dG1}
\frac{1}{\Gamma_{\tau}}\frac{d^2 \Gamma}{d M_{\text{inv}}(\pi^- f_0) d M_{\text{inv}}(\pi^+ \pi^-)}  = \frac{1}{(2 \pi)^5} \frac{1}{\Gamma_{\tau}} k \, p'_{\nu}\, \widetilde{q}_{\pi^+} \, \frac{2 m_\tau 2 m_\nu}{4 M^2_\tau}
\overline{\sum} \sum \left|t\right|^2 \, ,
\end{eqnarray}
with $k$ given by Eq.\eqref{eq:25} and
\begin{eqnarray}
p'_{\nu} = \frac{\lambda^{1/2} (m_{\tau}^2,m^2_\nu, M^2_{\text{inv}}(\pi^- f_0))}{2 m_{\tau} },	\quad
\widetilde{q}_{\pi^+}=\frac{\lambda^{1/2} ( M^2_{\text{inv}}(\pi^+ \pi^-),m^2_{\pi},m^2_{\pi})}{2 M_{\text{inv}}(\pi^+ \pi^-)}.
\end{eqnarray}

Similarly, for the $\tau^- \to \nu_\tau \pi^- \pi^0 \eta $ decay,  we can get the  double  differential  mass distribution for $M_{\rm inv}(\pi^0 \eta)$ and  $M_{\rm inv}(\pi^- a_0)$.

Note that the term ${m_\tau m_\nu}$ in the numerator of Eq.\eqref{eq:dG1} cancels the same factor in the denominator of Eqs. \eqref{eq:tf0} and \eqref{eq:ta0}.
In Eq.\eqref{eq:dG1} we have the factor $\frac{{\cal C}^2}{\Gamma_{\tau}}$, which, as mentioned before,  is obtained by means of  Eq.\eqref{eq:C}, and thus we can provide absolute values for the mass
distributions.

\section{Results}

\begin{figure}[ht!]
\includegraphics[scale=1.1]{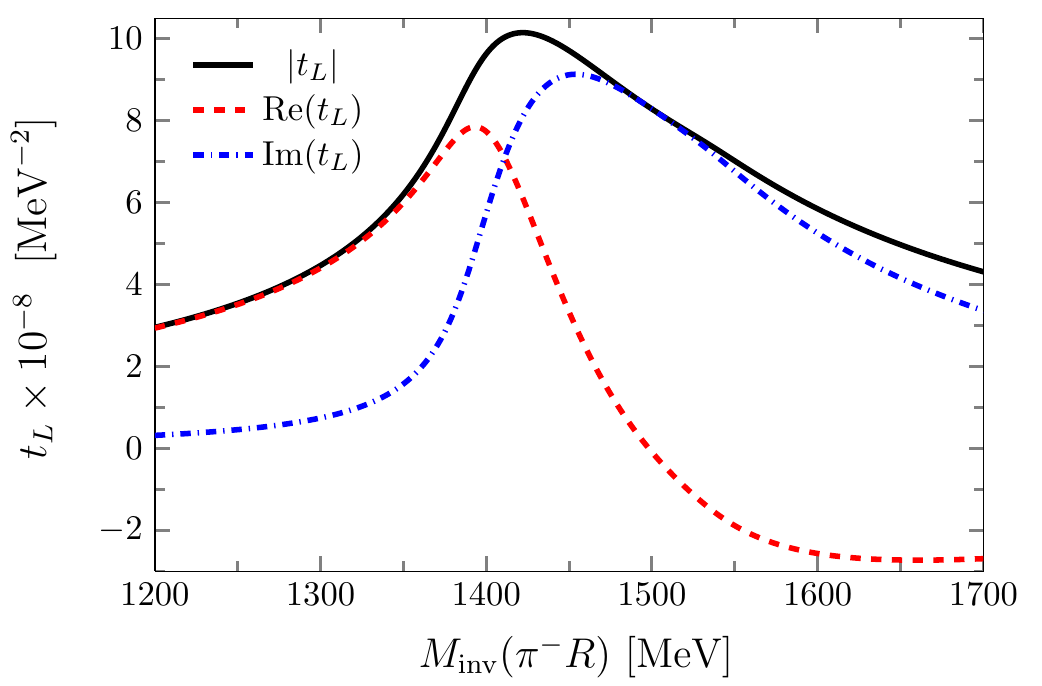}\\
\caption{Triangle amplitude  ${\rm Re} (t_L)$, ${\rm Im}(t_L)$ and $|t_L|$, taking $M_{\rm inv}(R)$=985 MeV     }
\label{fig:ttL}
\end{figure}

Let us begin by showing in Fig. \ref{fig:ttL} the contribution of the triangle loop defined in Eq. \eqref{eq:tt2}.
We plot the real and imaginary  parts of $t_L$, as well as the absolute value as a function of $M_{\rm inv}(\pi^- R)$,  with $M_{\rm inv}(R)$ fixed at $985$ MeV ($R$ standing for  $f_0(980)$ or $a_0(980)$).
It can be observed that ${\rm Re}(t_T)$  has a peak around $1393$ MeV, and  ${\rm Im}(t_T)$  has a peak around $1454$ MeV,
and  there is a peak for $|t_T|$ around 1425 MeV. As discussed in Refs.~\cite{sakairamos,dailam}, the peak of the real part is
related to the $K^* K$ threshold and the one of the imaginary part,  that dominates for the larger $\pi^- R$ invariant masses, to the triangle singularity.
Note that around $1420$ MeV and above the triangle singularity dominates the reaction.

The  origin of the peak in $|t_T|$  and consequently  in the $\pi^- R$ mass distribution of the decay has then the same origin as the peak observed in the COMPASS experiment \cite{compass},
tentatively branded as a new ``$a_1(1420)$" resonance, which however was explained in \cite{ketzer,aceti} as coming from the same triangle  mechanism that we have encountered here.
It would be most enlightening to confirm this experimentally in the $\tau$ decay reaction  to settle discussions around the ``$a_1(1420)$"  peak.

\begin{figure}[ht]
\includegraphics[scale=0.84]{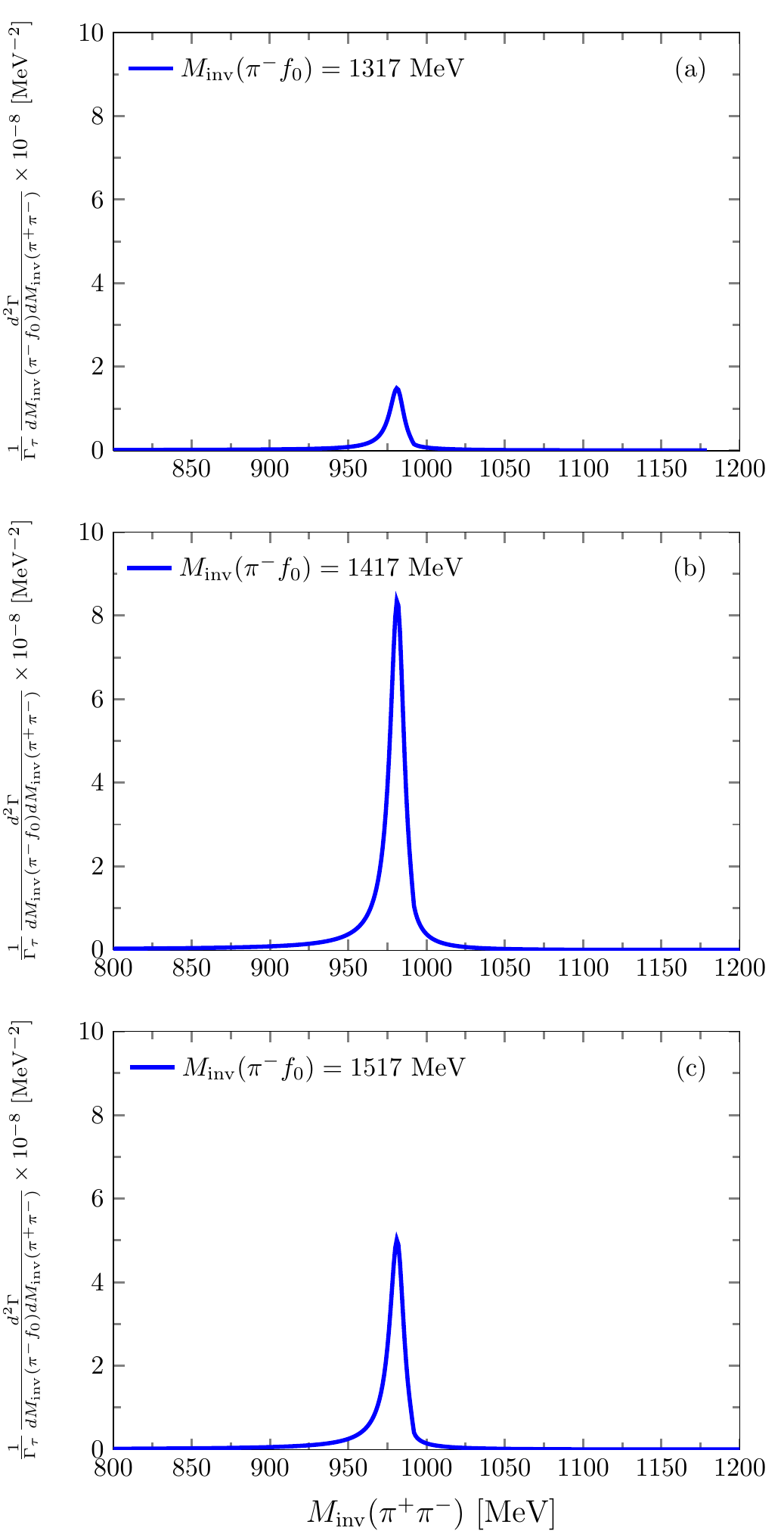}
\caption{Double differential width of  $\tau^- \to \nu_\tau \pi^- \pi^+ \pi^- $, keeping $M_{\rm inv}(\pi^- f_0)$  fixed to three values.
Lines (a),(b) and (c) show the values at $M_{\rm inv}(\pi^- f_0)$  1317 MeV, 1417 MeV, and 1517 MeV, respectively, plotted versus  $M_{\rm inv}(\pi^+ \pi^-)$. }
\label{fig:f0}
\end{figure}

\begin{figure}[ht]
\includegraphics[scale=0.84]{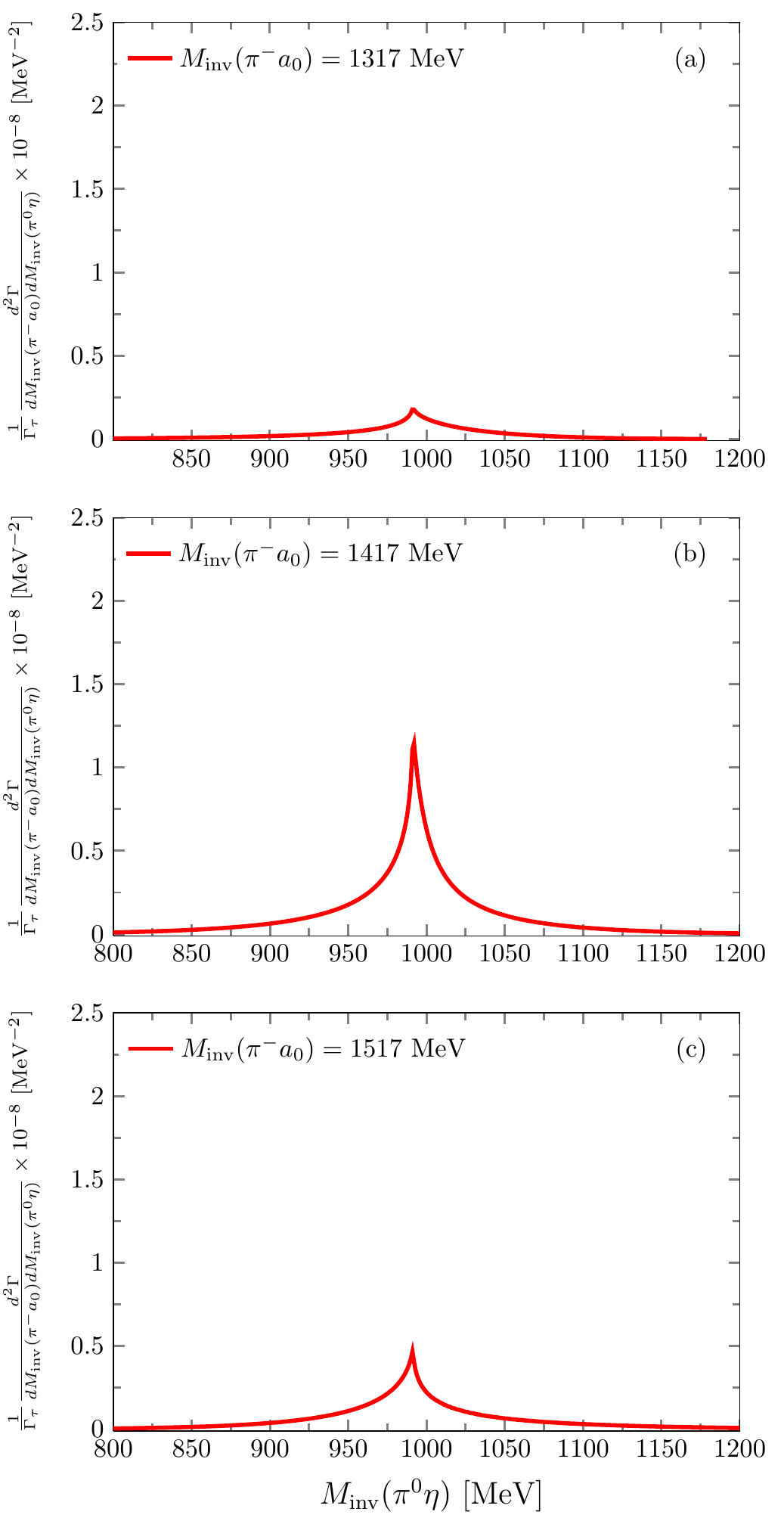}
\caption{Double differential width of  $\tau^- \to \nu_\tau \pi^- \pi^0 \eta $, keeping $M_{\rm inv}(\pi^- a_0)$
fixed to three values. Lines (a),(b) and (c) show the values at $M_{\rm inv}(\pi^- a_0)$  1317 MeV, 1417 MeV, and 1517 MeV, respectively, plotted versus  $M_{\rm inv}(\pi^0 \eta)$. }
\label{fig:a0}
\end{figure}

In Fig.~\ref{fig:f0}  we plot Eq.\eqref{eq:dG1} for the $\tau^- \to \nu_\tau \pi^- \pi^+ \pi^- $  decay, and similarly in  Fig.\ref{fig:a0} for the
$\tau^- \to \nu_\tau \pi^- \pi^0 \eta $ decay as a function of $M_{\rm inv}(R)$, where in both figures we  fix $M_{\rm inv}(\pi^- R)$=1317 MeV, 1417 MeV, and 1517 MeV
and vary  $M_{\rm inv}(R)$.  We can see that the distribution with largest strength is near $M_{\rm inv}(\pi^- R)$=1417 MeV. In Fig.\ref{fig:f0}
we can also see a strong peak in the $\pi^+ \pi^-$ mass distribution  around  $980$ MeV for the three different masses of $M_{\rm inv}(\pi^- R)$, corresponding to the $f_0(980)$.
  Similarly, in Fig.\ref{fig:a0}  we  see the distinctive cusp like
$a_0(980)$ peak around  $990$ MeV for the $\pi^0 \eta$  mass distribution.
Consequently, we see that most of the contribution to the width $\Gamma$ comes from $M_{\rm inv}(R)=M_R$ (the nominal mass of the $f_0$ or $a_0$ resonance), and  we have strong contributions for $M_{\rm inv}(\pi^+ \pi^-) \in [950~\rm{MeV},1000~\rm{MeV}]$
and  $M_{\rm inv}(\pi^0 \eta) \in [900~\rm{MeV},1050~\rm{MeV}]$. Therefore,  when we
calculate the mass distribution $\frac{d \Gamma}{d M_{\rm inv}(\pi^- R)}$, we restrict the integral  to the limits already mentioned.

\begin{figure}[ht]
\includegraphics[scale=1.1]{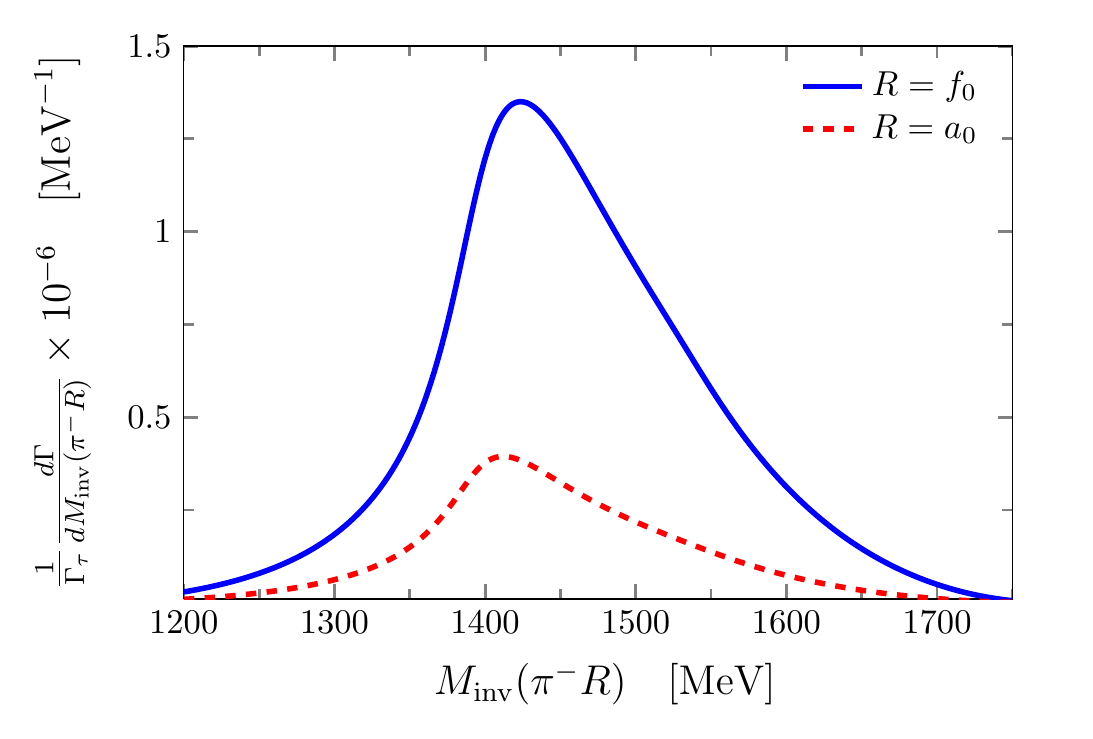}
\caption{The mass distribution for $\pi^- R$ ($R=f_0,a_0$). The solid line for  $\tau^- \to \nu_\tau \pi^- \pi^+ \pi^- $ as a function of $M_{\rm inv}(\pi^- R)$ with
$R \equiv f_0(980)$ measured in the $\pi^+ \pi^-$ decay mode;
dashed line for $\tau^- \to \nu_\tau \pi^- \pi^0 \eta $ as a function of $M_{\rm inv}(\pi^- R)$ with  $R \equiv a_0(980)$ measured in the $\pi^0 \eta$ decay mode. }
\label{fig:piR}
\end{figure}

By integrating over $M_{\rm inv}(R)$, we obtain $\frac{1}{\Gamma_{\tau}}\frac{d \Gamma}{d M_{\rm inv}(\pi^- R)}$ which is  shown in Fig.~\ref{fig:piR}.
We see a clear peak of the distribution around $1423$ MeV for $\pi^- f_0 (980)$ production and  $1412$ MeV for $\pi^- a_0 (980)$ production.
Integrating $\frac{d \Gamma}{d M_{\rm inv}(\pi^- R)}$ over  $M_{\rm inv}(\pi^- R)$ in Fig.~\ref{fig:piR}, we obtain the branching fractions
\begin{eqnarray}
{\cal B} (\tau \to \nu_\tau \pi^-  f_0(980); ~~ f_0(980)  \rightarrow \pi^+ \pi^-) &=& ( 2.6  \pm  0.5) \times 10^{-4}  \nonumber \,, \\
{\cal B} (\tau \to \nu_\tau \pi^-  a_0(980); ~~ a_0(980)  \rightarrow \pi^0 \eta)  &=& ( 7.1  \pm  1.4) \times 10^{-5}\, .
 \end{eqnarray}
 Since the rate of $f_0 \to \pi^0 \pi^0$ is one half that of $f_0 \to \pi^+ \pi^-$, we can write
 \begin{eqnarray}
{\cal B} (\tau \to \nu_\tau \pi^-  f_0(980); ~~ f_0(980)  \rightarrow \pi^+ \pi^-) &=& ( 3.9  \pm  0.8) \times 10^{-4} \, .
 \end{eqnarray}
 The errors in these numbers count only the relative error of the branching ratio of Eq. \eqref{eq:Ce}.
 These  numbers are within measurable range, since  branching ratios of $10^{-5}$ and smaller are quoted in the PDG for $\tau$ decays \cite{pdg}.

\section{Conclusions}
\label{sec:conc}
We have made a study of the $\tau^- \to  \nu_\tau \pi^- f_0(980)$  and $\tau^- \to  \nu_\tau \pi^- a_0(980)$ reactions from the perspective that the $f_0(980)$  and $a_0(980)$
are dynamically generated resonances  from the interaction of pseudoscalar mesons in coupled channnels. We showed that the formalism  for these processes
proceeds via $\tau \to \nu_\tau K^{*0} K^-$ ($ K^0 K^{*-}$) followed by $\bar{K}^{*} \to \pi^- K$  and the posterior fusion of $K\bar{K}$  to produce either the $f_0(980)$  or $a_0(980)$ states.
 This triangle mechanism  has a peculiarity since  it develops a triangle singularity at $M_{\rm inv} (\pi^- R) \simeq 1420$ MeV ($ R \equiv f_0$ or $a_0$), and the
 $M_{\rm inv} (\pi^- R)$ distribution  shows a peak around this energy, which has then the same origin as the explanations given in \cite{ketzer,aceti} for the COMPASS peak in $\pi f_0(980)$ that was initially
 presented  as the  new resonance ``$a_1(1420)$". It would be most instructive to have the experiment performed  to see if such peak indeed  appears, which would help clarify the issue
 around the  ``$a_1(1420)$" peak.

 On the other hand we make predictions which are tied to the way the $f_0(980)$  and $a_0(980)$ resonances are generated and again the observations will bring extra information on the nature of these low-lying
 scalar states.

 The mechanism requires the use of the amplitude  for $\tau \to \nu_\tau K^{*0} K^-$ reaction in a way suited to  the calculation of the loop function of the  triangle mechanism. This task was made
 efficient and easily manageable  thanks to the formalism  developed in \cite{tdai} which provides  two amplitudes with given $G$-parity in terms of the third components of
 the $K^{*0}$ spin. Since $\pi^- f_0(980)$ and   $\pi^- a_0(980)$  have  negative  and positive $G$-parity respectively, the formalism filtered  just one of these amplitudes for either reaction,
 with  the subsequent economy  and clarity in the formulation.

  We could provide absolute values for the mass distributions and final branching ratios by using the experimental branching ratio of the $\tau \to \nu_\tau K^{*0} K^-$ reaction. Hence, our predictions
  are free of intrinsic uncertainties  that ab initio  microscopic models unavoidably  have, and which would be magnified in this problem  where final state interaction of hadrons is at work.

With the reliable predictions of our approach we find final  branching ratios of $\pi^- f_0(980)$ and $\pi^- a_0(980)$ of about  $4 \times 10^{-4}$
 and $7 \times 10^{-5}$, respectively. These rates are well within measurable range and we can only encourage the performance of the experiments.
 Actually some partial information already exists for the reactions, exposed in \cite{inami} where the reaction $\tau^- \to  \nu_\tau \pi^-  \pi^0 \eta$ is measured, with a branching ratio  $1.38 \times 10^{-3}$, but
 the $\pi^0 \eta \to a_0(980)$ mode is not isolated, In \cite{buskulic} this reaction is also  measured and a peak seems to be present around
 $M_{\rm inv} (\pi^-  \pi^0 \eta) \simeq 1420$ MeV, but since the $\pi^0 \eta \to a_0(980)$  channel is not isolated we can not conclude  that this corresponds to
$\pi^- a_0(980)$. The same can be said about the work of \cite{artuso} where a peak around $1450$ MeV seems to be present in the $\pi^-  \pi^0 \eta$ invariant mass. As for
$\pi^-  \pi^+ \pi^-$  there are also studies in \cite{aubert,lee,briere} but no mass distributions are available. If the idea of building a $\tau$ facility in China prospers, the suggestion
of new decay modes and predictions  like those  in the present work will be most opportune to make such facility really useful. Meanwhile, the experiments just quoted, with larger statistic,
could produce new results to test our predictions.

\section*{Acknowledgments}

LRD acknowledges the support from the National Natural Science Foundation of China (Grant No. 11575076) and the State Scholarship Fund of China (No. 201708210057).
QXY  acknowledges the support from the National Natural Science Foundation of China (Grant Nos. 11775024 and 11575023).
This work is partly supported by the Spanish Ministerio de Economia y Competitividad and European FEDER funds under Contracts No. FIS2017-84038-C2-1-P B
and No. FIS2017-84038-C2-2-P B, and the Generalitat Valenciana in the program Prometeo II-2014/068, and
the project Severo Ochoa of IFIC, SEV-2014-0398 (EO).


\begin{thebibliography}{99}

\bibitem{landau}
  L.~D.~Landau,   Nucl.\ Phys.\  {\bf 13}, 181 (1959).
  %``On analytic properties of vertex parts in quantum field theory,''

\bibitem{coleman}
  S.~Coleman and R.~E.~Norton,   Nuovo Cim.\  {\bf 38}, 438  (1965).
 %``Singularities in the physical region,''

\bibitem{ketzer}
  M.~Mikhasenko, B.~Ketzer and A.~Sarantsev,   Phys.\ Rev.\ D {\bf 91}, 094015  (2015).
  %``Nature of the $a_1(1420)$,''

\bibitem{aceti}
  F.~Aceti, L.~R.~Dai and E.~Oset,   Phys.\ Rev.\ D {\bf 94},  096015 (2016).
  %``$a_1(1420)$ peak as the $\pi f_0(980)$ decay mode of the $a_1(1260)$,''

\bibitem{liu}
  X.~H.~Liu, M.~Oka and Q.~Zhao,    Phys.\ Lett.\ B {\bf 753}, 297 (2016).
  %``Searching for observable effects induced by anomalous triangle singularities,''

\bibitem{compass}
C. Adolph et al. (COMPASS Collaboration), Observation of
a New Narrow Axial-Vector Meson $a_1(1420)$, Phys. Rev. Lett. 115, 082001 (2015).


\bibitem{pdg}
M. Tanabashi et al. (Particle Data Group), Phys. Rev. D {\bf 98}, 030001 (2018).

\bibitem{barberis}
  D.~Barberis {\it et al.} [WA102 Collaboration],   Phys.\ Lett.\ B {\bf 440}, 225 (1998).
  %``A Measurement of the branching fractions of the f(1)(1285) and f(1)(1420) produced in central p p interactions at 450-GeV/c,''
%  doi:10.1016/S0370-2693(98)01264-7

 \bibitem{debastiani}
  V.~R.~Debastiani, F.~Aceti, W.~H.~Liang and E.~Oset,  Phys.\ Rev.\ D {\bf 95}, 034015 (2017).
  %``Revising the $f_1(1420)$ resonance,''

\bibitem{xie}
  J.~J.~Xie, L.~S.~Geng and E.~Oset,   Phys.\ Rev.\ D {\bf 95}, 034004 (2017).
  %``$f_2$(1810) as a triangle singularity,''


\bibitem{dailam}
  L.~R.~Dai, R.~Pavao, S.~Sakai and E.~Oset,    Phys.\ Rev.\ D {\bf 97}, 116004 (2018).


\bibitem{szczepaniak}
  A.~P.~Szczepaniak,  Phys.\ Lett.\ B {\bf 747}, 410 (2015).

\bibitem{szczepaniak2}
  A.~P.~Szczepaniak,  Phys.\ Lett.\ B {\bf 757}, 61 (2016).

\bibitem{bondar}
  A.~E.~Bondar and M.~B.~Voloshin,  Phys.\ Rev.\ D {\bf 93} 094008, (2016).

\bibitem{pilloni}
  A.~Pilloni {\it et al.} [JPAC Collaboration],  Phys.\ Lett.\ B {\bf 772}, 200  (2017).

\bibitem{pavao}
  R.~Pavao, S.~Sakai and E.~Oset,   Eur.\ Phys.\ J.\ C {\bf 77},  599 (2017).

%\bibitem{pasaos}
%  R.~Pavao, S.~Sakai and E.~Oset,   Eur.\ Phys.\ J.\ C {\bf 77}, 599  (2017).
%  %``Triangle singularities in $B^-\rightarrow D^{*0}\pi ^-\pi ^0\eta $ and $B^-\rightarrow D^{*0}\pi ^-\pi ^+\pi ^-$,''


\bibitem{liu2}
  X.~H.~Liu and U.~G.~Mei\ss ner,   Eur.\ Phys.\ J.\ C {\bf 77},  816 (2017).

%\bibitem{sakai}
%  S.~Sakai, E.~Oset and A.~Ramos,   Eur.\ Phys.\ J.\ A {\bf 54}, 10 (2018).

\bibitem{sakairamos}
  S.~Sakai, E.~Oset and A.~Ramos,   Eur.\ Phys.\ J.\ A {\bf 54}, 10  (2018).
%``Triangle singularities in $B^-\rightarrow K^-\pi^-D_{s0}^+$ and $B^-\rightarrow K^-\pi^-D_{s1}^+$,''


  \bibitem{rocaprc}
  L.~Roca and E.~Oset,   Phys.\ Rev.\ C {\bf 95},  065211 (2017).
  %``Role of a triangle singularity in the $\pi \Delta$ decay of the $N(1700)(3/2^-)$,''

\bibitem{daris}
   D.~Samart, W.~H.~Liang and E.~Oset,   Phys.\ Rev.\ C {\bf 96}, 035202 (2017).
  %``Triangle mechanisms in the build up and decay of the $N^*(1875)$,''


\bibitem{wuzou}
  J.~J.~Wu, X.~H.~Liu, Q.~Zhao and B.~S.~Zou,
  %``The Puzzle of anomalously large isospin violations in $\eta(1405/1475)\to 3\pi$,''
  Phys.\ Rev.\ Lett.\  {\bf 108}, 081803 (2012).
 % doi:10.1103/PhysRevLett.108.081803

\bibitem{acetiwu}
  F.~Aceti, W.~H.~Liang, E.~Oset, J.~J.~Wu and B.~S.~Zou,
  %``Isospin breaking and $f_0(980)$-$a_0(980)$ mixing in the $\eta(1405) \to \pi^{0} f_0(980)$ reaction,''
  Phys.\ Rev.\ D {\bf 86}, 114007 (2012).
  %doi:10.1103/PhysRevD.86.114007

\bibitem{wuwu}
  X.~G.~Wu, J.~J.~Wu, Q.~Zhao and B.~S.~Zou,
  %``Understanding the property of $\eta(1405/1475)$ in the $J/\psi$ radiative decay,''
  Phys.\ Rev.\ D {\bf 87},  014023 (2013).
%  doi:10.1103/PhysRevD.87.014023

\bibitem{liuli}
  X.~H.~Liu and G.~Li,
  %``Could the observation of X(5568) be a result of the near threshold rescattering effects?,''
  Eur.\ Phys.\ J.\ C {\bf 76}, 455 (2016).
 % doi:10.1140/epjc/s10052-016-4308-1


 \bibitem{ewang}
  E.~Wang, J.~J.~Xie, W.~H.~Liang, F.~K.~Guo and E.~Oset,
  %``Role of a triangle singularity in the $\gamma p\rightarrow K^+ \Lambda(1405)$ reaction,''
  Phys.\ Rev.\ C {\bf 95},  015205 (2017).
 % doi:10.1103/PhysRevC.95.015205

\bibitem{xieguo}
  J.~J.~Xie and F.~K.~Guo,
  %``Triangular singularity and a possible $\phi p$ resonance in the $\Lambda^+_c \to \pi^0 \phi p$ decay,''
  Phys.\ Lett.\ B {\bf 774}, 108 (2017).
%  doi:10.1016/j.physletb.2017.09.060


\bibitem{caozhao}
Z. Cao, Q. Zhao,    arXiv:1711.07309 [hep-ph].


\bibitem{liangsa}
  W.~H.~Liang, S.~Sakai, J.~J.~Xie and E.~Oset,
  %``Triangle singularity enhancing isospin violation in $\bar B_s^0 \to J/\psi \pi^0 f_0(980)$,''
  Chin.\ Phys.\ C {\bf 42},  044101  (2018).
 % doi:10.1088/1674-1137/42/4/044101


\bibitem{desaos}
  V.~R.~Debastiani, S.~Sakai and E.~Oset,
  %``Considerations on the Schmid theorem for triangle singularities,''
  arXiv:1809.06890 [hep-ph].


\bibitem{tdai}
 L.~R.~Dai, R.~Pavao, S.~Sakai and E.~Oset,   arXiv:1805.04573 [hep-ph].


\bibitem{roca}
 E.~Oset and L.~Roca,  Phys.\ Lett.\ B {\bf 782}, 332  (2018).
  %``Triangle singularity in $\tau \to f_1(1285)\pi\nu_\tau$ decay,''

\bibitem{roca2}
  L.~Roca, E.~Oset and J.~Singh,   Phys.\ Rev.\ D {\bf 72} 014002, (2005).
  %``Low lying axial-vector mesons as dynamically generated resonances,''
%  doi:10.1103/PhysRevD.72.014002

\bibitem{geng}
  Y.~Zhou, X.~L.~Ren, H.~X.~Chen and L.~S.~Geng,   Phys.\ Rev.\ D {\bf 90},  014020 (2014).
  %``Pseudoscalar meson and vector meson interactions and dynamically generated axial-vector mesons,''
  %  doi:10.1103/PhysRevD.90.014020

\bibitem{volkov}
  M.~K.~Volkov, A.~A.~Pivovarov and A.~A.~Osipov,   Eur.\ Phys.\ J.\ A {\bf 54},  61  (2018).
  %``$\tau \rightarrow f_{1}(1285) \pi^{-}\nu_{\tau}$ decay in the extended Nambu-Jona-Lasinio model,''

\bibitem{oller}
  J.~A.~Oller and E.~Oset,
  %``Chiral symmetry amplitudes in the S wave isoscalar and isovector channels and the $\sigma$, f$_0$(980), a$_0$(980) scalar mesons,''
  Nucl.\ Phys.\ A {\bf 620}, 438  (1997);  A {\bf 652}, 407 (E) (1999).

\bibitem{nieves}
  J.~Nieves and E.~Ruiz Arriola,   Nucl.\ Phys.\ A {\bf 679}, 57 (2000).

\bibitem{kaiser}
  N.~Kaiser,   Eur.\ Phys.\ J.\ A {\bf 3}, 307 (1998).

\bibitem{daiplb} J.~J.~Xie, L.~R.~Dai and E.~Oset,    Phys.\ Lett.\ B {\bf 742}, 363 (2015).

\bibitem{locher}
  M.~P.~Locher, V.~E.~Markushin and H.~Q.~Zheng,
  %``Structure of f0 (980) from a coupled channel analysis of S wave pi pi scattering,''
  Eur.\ Phys.\ J.\ C {\bf 4}, 317  (1998).
%  doi:10.1007/s100529800766, 10.1007/s100520050210

\bibitem{micu}
L. Micu, Nucl. Phys. B {\bf 10}, 521 (1969).

\bibitem{oliver}
  A.~Le Yaouanc, L.~Oliver, O. P\`{e}ne and J.~C.~Raynal, Phys. Rev. D {\bf  8}, 2223 (1973).

\bibitem{bijker}
  E.~Santopinto and R.~Bijker,   Phys.\ Rev.\ C {\bf 82}, 062202 (2010).
  %``Flavor asymmetry of sea quarks in the unquenched quark model,''
%  doi:10.1103/PhysRevC.82.062202

\bibitem{pr88}
B. C. Barish,  R. Stroynowski, Phys.  Rept.   {\bf 157}, 1  (1988).

\bibitem{mandl}
 F. Mandl and G. Shaw, Quantum Field Theory, John Wiley \& Sons, 1984.


\bibitem{aceti2}
F. Aceti, J. M. Dias and E. Oset,   Eur.\ Phys.\ J.\ A {\bf 51}, 48  (2015).

\bibitem{guo}
  M.~Bayar, F.~Aceti, F.~K.~Guo and E.~Oset,   Phys.\ Rev.\ D {\bf 94}, 074039  (2016).


\bibitem{inami}
  K.~Inami {\it et al.} [Belle Collaboration],   Phys.\ Lett.\ B {\bf 672},  209 (2009).
  %``Precise measurement of hadronic tau-decays with an eta meson,''

\bibitem{buskulic}
  D.~Buskulic {\it et al.} [ALEPH Collaboration],  Z.\ Phys.\ C {\bf 74},  263  (1997).
  %``A Study of tau decays involving eta and omega mesons,''

\bibitem{artuso}
 M.~Artuso {\it et al.} [CLEO Collaboration],   Phys.\ Rev.\ Lett.\  {\bf 69}, 3278 (1992).
  %``Measurement of tau decays involving eta mesons,''

\bibitem{aubert}
  B.~Aubert {\it et al.} [BaBar Collaboration],   Phys.\ Rev.\ Lett.\  {\bf 100}, 011801 (2008).

\bibitem{lee}
  M.~J.~Lee {\it et al.} [Belle Collaboration],   Phys.\ Rev.\ D {\bf 81}, 113007 (2010).

\bibitem{briere}
  R.~A.~Briere {\it et al.} [CLEO Collaboration],   Phys.\ Rev.\ Lett.\  {\bf 90}, 181802  (2003).

\end{thebibliography}
\end{document}